**Analysis of airport runway pavement reliability considering temperature variation: The case of São Paulo-Congonhas international airport**

Felipe H. Cava, M. Sc.,[1] Dimas B. Ribeiro, D. Sc.,[2] Claudia A. Pereira, D. Sc.,[3] Mauro Caetano, D. Sc.[4] and Evandro José da Silva, D. Sc.[5]

[1] Researcher, Civil Engineering Division, Aeronautics Institute of Technology, São José dos Campos, Brazil, 12228-900; email: felipe.cava.101745@ga.ita.br (Corresponding author)

[2] Researcher, Civil Engineering Division, Aeronautics Institute of Technology, São José dos Campos, Brazil, 12228-900; email: dimas@ita.br

[3] Researcher, Civil Engineering Division, Aeronautics Institute of Technology, São José dos Campos, Brazil, 12228-900; email: claudia.azevedo@ita.br

[4] Researcher, Air Transport Laboratory, Aeronautics Institute of Technology, São José dos Campos, Brazil, 12228-900; email: caetano@ita.br

[5] Researcher, Civil Engineering Division, Aeronautics Institute of Technology, São José dos Campos, Brazil, 12228-900; email: evandro@ita.br

**ABSTRACT:**

Airport pavement design methods typically rely on standard documents, such as those provided by the FAA, which assume general climatic conditions. Nonetheless, the temperature between different regions tends to influence the behavior of the pavements, which impacts how stress and strains are distributed within the pavement structure and influence pavement performance and reliability. Furthermore, global warming has required specific analyses of the behavior of infrastructures, such as pavements, regarding the choice of materials and performance needed to make pavements more resilient. This study aims to perform a reliability analysis for a pavement designed by traditional methods combined with the temperature variation, considering the case of São Paulo-Congonhas International Airport (CGH). It analyzed temperature variations across the four seasons. The procedure includes designing pavement, considering the airport's traffic mix, and performing a Monte Carlo Simulation (MCS) to verify the pavement structure's reliability under temperature variation. The study shows that the total cumulative damage factor (CDF) computed through MCS is 79% lower than the value obtained using FAA method. Considering the pavement temperatures at CGH, all aircraft tend to cause less damage than expected. Furthermore, the pavement designed could withstand traffic 2.5 times greater at 95% reliability and 5.0 times greater at 50% reliability when considering temperature variation. These numbers indicate that in Brazilian airports where fatigue is the primary design criterion, FAARFIELD overestimates the damage and consequently increases pavement construction costs. These results suggest that the airport pavement design method requires calibration for Brazilian climatic conditions to improve fatigue damage prediction, especially for airports where fatigue is the primary failure criterion. The limitations of this study should be acknowledged to inform future research. The pavement temperature equation applied is deterministic, assuming fixed values for albedo, wind speed, and atmospheric transmission. Future research should assess the suitability of this equation for Brazilian regions, particularly in relation to actual measured temperatures at pavement depth. In this study, pavement reliability was evaluated considering only temperature variations; factors such as precipitation and variability in pavement thickness were not included, although they may affect pavement performance. Additionally, fatigue tests under different asphalt temperatures were not conducted, and a standard stiffness value for P-401 was used to assess fatigue behavior. Future studies by the authors will aim to calibrate the performance equations and address these limitations.

**INTRODUCTION:**

The reliability of airport pavements is important for ensuring safety. Some airports require particular attention due to the high volume of traffic and environmental conditions. Various climatic factors may influence pavement performance, such as surface energy balance, moisture, climate change, frost heaving, and temperature (Alavi, Pouranian and Hajj, 2014). Thus, among all these factors, temperature is considered the critical variable, directly affecting the material behavior and performance (Qiao et al., 2013; Cheng et al., 2020; Zhang et al., 2023).

In countries with continental dimensions, such as Brazil, the difference in temperature between the states to the north and south tend to influence the behavior of the pavements. These differences in asphalt layer temperature impact how stress and strain are distributed within the pavement structure and influence pavement performance and reliability (Hasan, Hiller and You, 2015; Kodipilly et al., 2018; Zhang et al., 2023; Luo et al., 2023; Zhuang et al., 2024).

Furthermore, global warming has required specific analyses of the behavior of infrastructures, such as pavements, regarding the choice of materials and performance needed to make pavements more resilient. Studies in several regions have been carried out to verify the impacts of climate change on asphalt pavements (Zhang et al., 2022; Liu et al, 2023; Barbi Tavassoti and Tighe, 2023; Zhang, Yang and Chen, 2024; Yang et al, 2024; Hosseini et al., 2024). All these studies agree that global warming tends to reduce pavement lifespan and increase pavement reconstruction costs.

Nonetheless, the FAA pavement design method considers a constant asphalt field temperature, which is assumed to be equal to 32°C over the design period (FAA, 2021). That is, the airport pavement design method does not differentiate the damage caused by aircraft in different weather seasons. Furthermore, the fatigue equation used in the pavement design method was not calibrated for field conditions (Shen and Carpenter, 2005; Shen and Carpenter, 2007), which tends to be problematic for airports where fatigue is the primary failure criterion.

In this scenario, the objective of this study is to perform a reliability analysis of the pavement designed using FAARFIELD software for a specific airport in Brazil and consider temperature variation. Climatic data from the past 10 years were collected to analyze the thermal gradient in the asphalt layers of the pavement structure. These data were also used to perform a sensitivity analysis on how temperature affects pavement strains to understand the impact of climatic conditions on pavements. Subsequently, a reliability analysis was conducted on CGH by comparing the results from FAARFIELD with those obtained from the MCS analysis. It was observed that the thermal gradient present in the pavement structure increases the stiffness of the asphalt layers, reducing the stresses and strains acting along the depth. Additionally, for the CGH, it was noted that the current pavement design method overestimates the stresses acting on runways.

**METHODS**

The methodology is divided into four parts.

In the first part, according to Figure 1, the collection and data analysis of traffic data and air temperature were made to São Paulo-Congonhas International Airport (CGH). This study chose this airport because it is one of the busiest in Brazil. Infraero provided the traffic data for 2017, 2018 and 2022, corresponding to every landing and takeoff performed by each aircraft on the airport's runway. Data were provided separately for each year and were then computed to determine each aircraft's average annual takeoffs and landings.

The air temperature was obtained from meteorological data provided by the BDMEP-INMET website. The application requires entering the time period and station type to retrieve meteorological data. This study utilized air temperature data from 2014 to 2024 from the Mirante de Santana meteorological station, which is the closest meteorological station to CGH. The temperature of the asphalt layer was obtained through the method developed by Huber (1994), which is also used in the SUPERPAVE software to select the asphalt binder according to the weather conditions. For this study, Equations 1 and 2 present the model considering CGH latitude.

$$T_{surf} = T_{air} + 15.5403 \tag{1}$$

$$T_d = \left[ T_{surf} \cdot \left( 1 - 0.0630 \cdot \frac{d}{25.4} + 0.0070 \cdot \left( \frac{d}{25.4} \right)^2 - 0.0004 \cdot \left( \frac{d}{25.4} \right)^3 \right) - 32 \right] \cdot \frac{5}{9} \tag{2}$$

Equation 1 corresponds to the temperature at the pavement surface $T_{surf}$ in CGH, calculated using air temperature $T_{air}$ in degrees Celsius. Equation 2 computes the pavement temperature profile $T_d$ using the layer depth $d$ in millimeters and the pavement temperature surface in degrees Celsius. These equations consider an albedo of 10%, transmission through air of 81%, atmospheric radiation of 70%, and 4.5 meters per second wind speed. The author developed them using more than 6,000 weather stations (Huber, 1994).

The second part of the method consists of the pavement design for the traffic data using FAARFIELD software. It designs the pavement considering the materials from the FAARFIELD bibliography (FAA, 2021) and uses them as a reference for the following analysis. After obtaining the pavement structure, a sensitivity analysis was performed in the third part, varying only the Hot Mix Asphalt (HMA) modulus based on the asphalt layer temperature. Equations 3 to 5 use laboratory tests conducted by Kuchiishi, Vasconcelos, and Bernucci (2019) on asphalt mixtures for this correlation.

$$\log \alpha_T = 0.00116 \cdot T^2 + 0.19600 \cdot T + 3.45 \tag{3}$$

$$f_r = f \cdot a_T \tag{4}$$

$$\log|E_{HMA}| = 1.690 + \frac{2.780}{1 + e^{-1.410 - 0.719 \cdot \log f_r}} \tag{5}$$

Equation 3 corresponds to the shift factor $\alpha_T$ of the model, calculated for each asphalt temperature T in degrees Celsius. Equation 4 calculates the reduced frequency $f_r$, representing the effect of frequency $f$ in the dynamic modulus test. This study considers the frequency of 1Hz, which is the same as computing the resilient modulus in Brazil. Equation 5 gives the HMA dynamic modulus $E_{HMA}$ in MPa. Equations 3 to 5 correspond to the dynamic modulus sigmoidal model, with coefficients obtained by Kuchiishi, Vasconcelos, and Bernucci (2019) for asphalt mixture samples.

The third part involves a sensitivity analysis that considers the traffic, climatic data, and pavement structure obtained in this method's first and second parts. The method performs a sensitivity analysis to obtain the pavement behavior under temperature variation and uses it in the last part.

Finally, this work employs a Monte Carlo Simulation (MCS) to analyze the reliability of pavement structures designed using FAARFIELD software for Brazilian climatic conditions. The MCS is a computerized technique that generates random values for the independent variable (Toan et al., 2022). Based on these values, the method obtained reliability $R$ by counting the number of variable combinations that satisfy the design criteria relative to the number of random variables generated, according to Equations 6 to 8. Many authors have used this technique to analyze pavement reliability (Maji & Das, 2008; Dilip & Babu, 2013; Ioannides & Tingle, 2021; Norouzi et al., 2022; Dinegdae, Ahmed and Erlingsson, 2022).

$$f(CDF > 1) = 1 \qquad (6)$$

$$f(CDF \leq 1) = 0 \qquad (7)$$

$$R = \frac{samples\ with\ f(CDF \leq 1)}{total\ number\ of\ samples} \qquad (8)$$

This study used the MCS method to generate values for asphalt temperature per season. A convergence analysis determined the number of random values adequate for the MCS. These values were then employed to calculate the tensile strains on pavement structure. Subsequently, the reliability was calculated considering the number of samples that met the design criteria for airport pavement structures. A Microsoft Excel® spreadsheet developed by the authors performed pavement reliability calculations.

The following section discusses the results of this analysis on enhancing airport pavement projects and maintenance.

**RESULTS**

This study divides the results into four parts: traffic and climatic data, pavement design, sensitivity analysis, and pavement reliability.

**Traffic and climatic data**

The traffic data analysis considered the most common aircraft at CGH. Note that B738 is the primary aircraft, performing an average of approximately 26,000 takeoffs and landings per year. Figure 2 illustrates the airport's leading aircraft, representing 90% of the traffic data. The remaining 10% consists of smaller aircraft that are not considered relevant for pavement design and analysis. Furthermore, the differences in operations over the years are less than 0.5%.

Figure 2 shows that the number of takeoffs and landings in CGH are similar. Only takeoffs were considered in the pavement design, as they represent the critical loading condition. About the climatic data, Figure 3 illustrates the dispersion of temperature data collected and shows that air temperature presents a seasonal variation, decreasing and increasing over time. Thus, data were analyzed considering the probability density of air temperature for each season in the Southern Hemisphere, according to Figure 4.

According to Figure 4, January to March typically exhibits the highest temperature, corresponding to summer in Brazil, while winter generally experiences the lowest temperature. Furthermore, winter is the season with the most significant coefficient of variation (COV), i.e., there is more air temperature dispersion among this season's months. These differences in air temperature and pavement temperature also change the HMA modulus due to the viscoelastic behavior of asphalt mixtures.

For reliability purposes, the Kolmogorov-Smirnov test was performed on air temperature data per season to verify normality. Figure 5 illustrates the histogram of air temperature per season.

According to Figure 5 and the Kolmogorov-Smirnov test, temperature follows a normal distribution with a p-value of 0.8044 in the worst case. Then, the reliability analysis conducted in this study considers a normal temperature distribution for each season.

**Pavement design using FAARFIELD**

The airport pavement was designed to achieve a minor pavement structure capable of sustaining the projected traffic load for 20 years. Table 1 presents the pavement structure obtained and each layer's modulus (E), corresponding to a CDF of 1.00 and 0.00 for fatigue and rutting, respectively. CDF corresponds to the ratio between the actual number of load applications and the allowable number of coverages for a given failure model, and it is a dimensionless parameter.

According to Table 1, the pavement structure requires a total thickness of 1,070 millimeters, with 250 millimeters of asphalt. The pavement design method considers the HMA modulus equal to 1,378.95MPa, obtained for 32°C according to the FAA (2021). Table 2 presents the CDF for each aircraft.

Comparing the CDF contribution for fatigue and subgrade leads to the conclusion that fatigue is the critical failure criterion for pavement design at CGH. The type of aircraft operating at the airport explains this conclusion, as they have similar main gears, MTOW, and high-intensity takeoff, which cause the pavement to be consistently stressed in the same areas. The B738 is the aircraft that causes the most damage to the pavement, representing 43% of total damage. Thus, the E195 aircraft causes the most minor damage, representing just 1% of total damage during the design period.

Based on this structure presented in Table 1, the method discretizes the asphalt layer into ten parts and calculates the pavement temperature considering the center of each layer. A convergence analysis was conducted, varying the discretization of the asphalt layer between 2 and 15 parts. Based on this analysis, the discretization was defined to be 10 parts. Applying Equations 1 and 2, it is observed an average three degrees Celsius difference between the centers of the asphalt layers. The HMA modulus was then computed using Equations 3 – 5. Table 3 presents the result.

According to Table 3, the average HMA modulus changes more than six times between the center of asphalt layers one and ten. Considering the first asphalt layer in the fall and winter, the HMA modulus is similar to that considered by FAARFIELD for the material P-401. Nonetheless, as the thickness increases the HMA modulus also increases due to the temperature profile. This difference shows that considering only one HMA modulus for thick asphalt layers, as FAARFIELD does, is unsuitable.

Furthermore, Table 3 shows differences in HMA modulus across the seasons. According to Table 3, there is an average variation of 76% in the HMA modulus between winter and summer. Additionally, there are similarities between the average temperatures of summer and spring, and winter and fall.

**Sensitivity Analysis**

The study's next step was performing a sensitivity analysis, varying the temperature of the first pavement layer and considering a range from 22°C to 41°C. The upper value was defined based on the maximum temperature calculated at the center of the first layer at CGH. The lower value was defined based on the probability distribution function, ensuring that only 2.5% of the probability is less than this value. Equations 3-5 led to an HMA modulus variation from 5,636MPa to 455MPa, respectively. The procedure obtains pavement strains using mePADS software (Maina et al., 2008), which considers the layered elastic theory. Figure 6 illustrates the influence of pavement temperature on tensile strain at the bottom of the asphalt layer.

According to Figure 6, tensile strains increase by 122% on average between 22°C and 41°C, indicating that the variation in temperature increases the pavement strains. In Brazil, labs usually measure the HMA modulus at a temperature of 25°C. Considering the average pavement temperature in summer, i.e., around 37°C in the center of the first asphalt layer, the tensile strain at the bottom of the HMA layer ($\varepsilon_{ht}$) is, on average, 47% higher than that observed at 25°C for all aircraft. This demonstrates that the temperature used in Brazil for HMA modulus tests is not consistent with the average pavement temperatures, which could lead to errors in considering a modulus higher than the field conditions during the pavement design. When considering the FAA model (FAA, 2021), this difference in pavement strains results, on average, in a reduction of 92% in fatigue life if the pavement is designed considering 25°C.

With the reduction of HMA modulus due to the pavement temperature, the compressive strains at the top of the subgrade (εvc) also increase on average by 48% between 22°C and 41°C, according to Figure 7.

Rutting is not the main design criterion in this work, as the strains leading to fatigue are more critical, as previously presented. Therefore, temperature variation does not reduce the life of the pavement structure. Thus, designers should consider this variation for airports where rutting is the main failure criterion. The increment in compressive strain at the subgrade may reduce the pavement life for rutting criteria, as mentioned by Zhang et al. (2023).

As fatigue is the primary design criterion in this study, a regression analysis was performed based on data from Figure 6 for each aircraft used in the reliability analysis. Equation 9 shows the general form of regression.

$$\varepsilon_{ht\,i} = \beta_0 \cdot E_{HMA}^{\beta_1} \qquad (9)$$

Table 4 summarizes the regression coefficients for each aircraft, obtained using the least squares method.

According to Table 4 and Equation 9, the regression analysis resulted in high coefficients of determination ($R^2$) to express the tensile strains at the bottom of the asphalt layer. This equation generates tensile strains at the bottom of asphalt layers according to the HMA modulus.

**Reliability Analysis**

This study performs a reliability analysis considering the normal distribution of temperature for each season and its influence on the tensile strain at the bottom of the HMA layer. The design period was divided into the four seasons of the year, analyzing each season's contribution and aircraft's contribution. A convergence analysis concluded that 5,000 samples are sufficient for the MCS, as shown in Figure 8.

According to Figure 8, the MCS reliability converges into approximately 1,500 samples. The standard deviation of the MCS is less than 0.1% for more than 2000 samples. In this study, 5,000 samples were used for the reliability analysis to reduce the standard deviation (SD) of the random sample generation, which is approximately 0,09% for 5,000 samples. Table 5 presents the average damage per aircraft and season using MCS.

According to Table 5, due to the high temperatures and reduced HMA modulus in summer, aircraft tend to cause more damage to the pavement structure. In the same way, the aircraft tend to cause the most minor damage in winter. Thus, the total CDF computed through MCS is 79% lower than the value obtained using FAARFIELD software. Considering the pavement temperature at CGH, all aircraft tend to cause less damage than expected. This tendency occurs when considering the thermal differential of the asphalt layer, i.e., the temperature decreases with the depth of the asphalt layer. Due to this reduction in the pavement temperature, the bottom of the asphalt layer has a higher modulus and consequently experiences less tensile strains, which is responsible for fatigue damage. According to White (2018), fatigue cracking is rarely encountered in airport pavements, which may be caused by these findings. Barbi, Tavassoti, and Tighe (2023) observations corroborate this result, even for regions without a pavement freezing period.

Considering the damage level observed using MCS, the pavement designed using FAARFIELD would withstand traffic 2.5 times greater at 95% reliability and 5.0 times greater at 50% reliability, according to Figure 9.

Clearly, traffic levels five times higher are unrealistic compared to field observations, which may indicate that the fatigue model is not suitable for Brazilian climate conditions and requires a laboratory to field calibration. Due to the minor damage caused when considering Brazilian climatic conditions at CGH, the design period could be approximately 50 years for 95% reliability and 102 years for 50% reliability. These design periods are much longer than those typically considered in airport pavement projects, which can be corroborated with the statements of White (2018) that fatigue cracks are not common in these pavements. On the other hand, the fact that they are not common also indicates a failure in predicting the structural behavior, possibly increasing construction costs, which requires a laboratory-field calibration. Furthermore, the need to calibrate the fatigue equation used in the design method was also mentioned by Shen and Carpenter (2007).

These differences between the CDF from the FAA method and those calculated via MCS in this study indicate that the fatigue model requires calibration for different regions. The fixed pavement temperature of 32°C during all the design periods is unrealistic because it does not account for the temperature differences between seasons.

**SUMMARY AND CONCLUSIONS**

This study performed a pavement reliability analysis considering CGH, the second busiest airport in Brazil. Furthermore, a sensitivity analysis was performed, considering pavement temperature variation.

This study showed that an increase in temperature also increases pavement strains for both tensile at the bottom of the asphalt layer and compressive strain at the top of the subgrade. The reliability analysis showed that the pavement designed using FAARFIELD might result in an unrealistic design period for less than 95% reliability, which may not be suitable for accurately predicting asphalt fatigue life and could increase pavement construction costs.

These unrealistic values may occur due to the absence of calibration in the fatigue performance prediction model used in the design software, indicating the need to calibrate the fatigue performance model for field conditions. Furthermore, Brazil exhibits significant climatic differences from north to south, and depending on the region analyzed, the pavement may experience different strains, which are not considered in the airport pavement design method.

The limitations of this study should be acknowledged to inform future research. The equation used in this study for pavement temperature is deterministic, always considering the same albedo, wind speed, and transmission through air. Future studies must verify if this equation is adequate for Brazilian regions, especially considering actual temperature values at pavement depth. This study performs pavement reliability by considering only the temperature variations, and then precipitation and the variability in pavement thickness were not considered, which may influence the pavement performance. Furthermore, this study did not perform fatigue tests varying the asphalt temperature and considered the standard stiffness value for P-401 for fatigue performance.

In future studies, the authors will focus on calibrating the fatigue equation for Brazilian climatic conditions. Laboratory tests are currently being conducted to evaluate the performance of asphalt mixtures under different temperature variations. In the next phase, a full-scale airport pavement structure will be constructed and monitored across different seasons. A digital twin, combining finite element analysis (FEA) and machine learning, will then be used to support the calibration process.

Additionally, future studies can focus on verifying the temperature equation for Brazilian airports and incorporating actual temperature values at pavement depth. Furthermore, fatigue tests at different asphalt temperatures are recommended to verify fatigue behavior and enhance pavement design methods for Brazilian airports.

**DATA AVAILABILITY STATEMENT:**

Some or all data, models, or code that support the findings of this study are available from the corresponding author upon reasonable request.

**NOTATION LIST**

The following symbols are used in this paper:

$E$ = layer modulus, MPa

$E_{HMA}$ = asphalt layer modulus, MPa

$R$ = Reliability

$T$ = temperature, ºC

$T_{Air}$ = Air temperature, ºC

$T_d$ = Pavement temperature at a specific depth, ºC

$T_{surf}$ = Temperature at the pavement surface, ºC

$d$ = depth, mm

$f$ = test frequency, Hz

$f_r$ = reduced frequency, Hz

$\alpha_T$ = shift factor at the temperature T.

$\beta_0, \beta_1$ = Regression coefficients

$\varepsilon_{ht_i}$ = tensile strain for aircraft "i".

**REFERENCE:**

AASHTO 2008. "Mechanistic-empirical pavement design guide: A manual of practice". Washington.

Alavi, M.Z; Pouranian, M.R; Hajj, E.Y 2014. "Prediction of Asphalt Pavement Temperature Profile with Finite Control Volume Method". Transportation Research Record: Journal of the Transportation Research Board. DOI: 10.3141/2456-10.

Barbi, P. S. R.; Tavassoti, P.; Tighe, S 2023. "Enhanced Pavement Design and Analysis Framework to Improve the Resiliency of Flexible Airfield Pavements". Transportation Research Record: Journal of the Transportation Research Board. DOI: 10.1177/03611981231155909.

Cheng, H; Liu, J; Sun, L; Liu, L; Zhang, Y 2020. "Fatigue behaviors of asphalt mixture at different temperatures in four-point bending and indirect tensile fatigue tests". Construction and Building Materials. DOI: https://doi.org/10.1016/j.conbuildmat.2020.121675

Dilip, D.M; P; Babu, G.L.S 2013. "Methodology for Pavement Design Reliability and Back Analysis Using Markov Chain Monte Carlo Simulation". Journal of Transportation Engineering, ASCE. DOI: https://doi.org/10.1061/(ASCE)TE.1943-5436.0000455.

Dinegdae, T; Ahmed, A; Erlingsson, S 2022. "Toward a comprehensive pavement reliability analysis approach". Transportation Research Record: Journal of the Transportation Research Board. DOI: https://doi.org/10.1177/03611981231155179.

FAA 2021. "AC 150/5320 6G - Airport Pavement Design and Evaluation". Washington.

Hasan, M.R.M; Hiller, J.E; You, Z 2015. "Effects of mean annual temperature and mean annual precipitation on the performance of flexible pavement using ME design". International Journal of Pavement Engineering. DOI: https://doi.org/10.1080/10298436.2015.1019504.

Hosseini, F; Nasimifar, M; Sivaneswaran, N; Golalipour, A 2024. "Mutual impacts of changing climate and flexible pavement performance considering resilience and sustainable aspects". Elsevier: Journal of Cleaner Production. DOI: https://doi.org/10.1016/j.jclepro.2024.142594.

Huber, G.A 1994. "SHRP-A-648A – Weather database for the SUPERPAVE mix design system". Strategic Highway Research Program. Washington, DC.

Ioannides, A.M; Tingle, J.S 2021. "Monte Carlo Simulation for flexible pavement reliability". Airfield and Highway pavements, ASCE. DOI: 10.1061/9780784483503.002.

Kodippilly, S; Yeaman, J; Henning, T; Tighe, S 2018. "Effects of extreme climatic conditions on pavement response". Road Materials and Pavement Design. DOI: https://doi.org/10.1080/14680629.2018.1552620.

Kuchiishi, A. K; Vasconcelos, K; Bernucci, L.B 2019. "Effect of mixture composition on the mechanical behaviour of cold recycled asphalt mixtures". International Journal of Pavement Engineering. DOI: https://doi.org/10.1080/10298436.2019.1655564.

Liu, T; Yang, S; Jiang, X; Liao, B; Castillo-Camarena, E 2023. "Adaptation measures for asphalt pavements to climate change in China". Elsevier: Journal of Cleaner Production. DOI: https://doi.org/10.1016/j.jclepro.2023.137861.

Luo, Y; Wu, H; Song, W; Yin, J; Zhan, Y; Yu, J; Wada, S.A 2023. "Thermal fatigue and cracking behaviors of asphalt mixtures under different temperature variations". Construction and Building Materials. DOI: https://doi.org/10.1016/j.conbuildmat.2023.130623.

Maina, J.W., Denneman, E; De Beer, M. Introduction of new road pavement response modelling software by means of benchmarking. Partnership for research and progress in Transportation. 27th Southern African Transport Conference (SATC), Pretoria, South Africa, July 7-11, 2008, pp 1-14.

Maji, A; Das, A 2008. "Reliability considerations of bituminous pavement design by mechanistic-empirical approach". International Journal of Pavement Engineering. DOI: https://doi.org/10.1080/10298430600997240.

Norouzi, Y; Ghasemi, S.H; Nowak, A.S; Jalayer, M; Mehta, Y; Chmielewski, J 2022. "Performance-based design of asphalt pavements concerning the reliability analysis". Construction and Building Materials, Vol. 332, 16. DOI: https://doi.org/10.1016/j.conbuildmat.2022.127393

Qiao, Y; Flintsch, G; Dawson, A; Parry, T 2013. "Examining effects of climatic factors on flexible pavement performance and service life". Transportation Research Record: Journal of the Transportation Research Board, p. 100–107. DOI: 10.3141/2349-12.

Shen, S; Carpenter, S.H 2005. "Application of the Dissipated Energy Concept in Fatigue Endurance Limit Testing". Transportation Research Record: Journal of the Transportation Research Board. DOI: https://doi.org/10.1177/0361198105192900120.

Shen, S; Carpenter, S.H 2007. "Development of an asphalt fatigue model based on energy principles". Asphalt Paving Technology.

Toan, T.D; Long, N.H; Wong, Y.D; Nguyen, T 2022. "Effects of variability in thickness and elastic modulus on the reliability of flexible pavement structural performance". International Journal of Pavement Engineering, Taylor & Francis. DOI: https://doi.org/10.1080/10298436.2022.2039923.

White, Greg 2018. "State of the art: asphalt for airport pavement surfacing". International Journal of Pavement Research and Technology. DOI: https://doi.org/10.1016/j.ijprt.2017.07.008.

Yang, Q; Cao, Z; Shen, L; Gu, F; Santos, J; Qiao, Y; Wang, H; Li, J; Zhang, Y; Chu, C 2024. "Impacts of climate change on environmental and economic sustainability of flexible pavements across China". Elsevier: Resources, Conservation & Recycling. DOI: https://doi.org/10.1016/j.resconrec.2024.107589.

Zhang, C; Tan, Y; Gao, Y; Fu, Y; Li, J; Li, S; Zhou, X 2022. "Resilience assessment of asphalt pavement rutting under climate change". Elsevier: Transportation Research Part D. DOI: https://doi.org/10.1016/j.trd.2022.103395.

Zhang, K; Wang, S; Yang, W; Zhong, X; Liang, S; Tang, Z; Quan, W 2023. "Influence of temperature and humidity coupling on rutting deformation of asphalt pavement". Science and Engineering of composite materials. DOI: https://doi.org/10.1515/secm-2022-0232.

Zhang, Q; Yang, S; Chen, G 2024. "Regional variations of climate change impact on asphalt pavement rutting distress". Elsevier: Transportation Research Part D. DOI: https://doi.org/10.1016/j.trd.2023.103968.

Zhuang, C; Guo, H; Zhao, S; Shu, S; Ye, Y; Xing, B 2024. "Study on fatigue performance of asphalt mixture in service life based on accelerated loading test". Construction and Building Materials. DOI: https://doi.org/10.1016/j.cscm.2024.e03055.

**TABLES:**

**Table 1.** Pavement thickness and modulus

| Layer | FAARFIELD Material | Thickness (mm) | E (MPa) |
|---|---|---|---|
| HMA | P-401 / P403 | 250 | 1,378.95 |
| Crushed Aggregate | P-209 | 400 | 508.87 |
| Uncrushed Aggregate | P-154 | 420 | 148.92 |
| Soil | Subgrade | - | 103.42 |

**Table 2.** CDF Contribution for each aircraft

| Aircraft | Fatigue CDF Contribution | Subgrade CDF Contribution | P/C Ratio |
|---|---|---|---|
| B738 | 0.43 | 0.00 | 2.01 |
| A320 | 0.18 | 0.00 | 2.06 |
| A319 | 0.24 | 0.00 | 2.04 |
| B737 | 0.14 | 0.00 | 2.04 |
| E195 | 0.01 | 0.00 | 2.12 |

**Table 3.** HMA modulus for each season and layers

| Asphalt Layer | Variable | Season | | | |
|---|---|---|---|---|---|
| | | Summer | Fall | Winter | Spring |
| 1 | Average Pavement Temperature (°C) | 37 | 33 | 33 | 36 |
| | HMA modulus (MPa) | 720 | 1217 | 1217 | 817 |
| 2 | Average Pavement Temperature (°C) | 35 | 30 | 30 | 33 |
| | HMA modulus (MPa) | 930 | 1855 | 1855 | 1217 |
| 3 | Average Pavement Temperature (°C) | 32 | 28 | 28 | 31 |
| | HMA modulus (MPa) | 1398 | 2470 | 2470 | 1609 |
| 4 | Average Pavement Temperature (°C) | 31 | 27 | 26 | 29 |
| | HMA modulus (MPa) | 1609 | 2849 | 3282 | 2140 |
| 5 | Average Pavement Temperature (°C) | 29 | 25 | 25 | 28 |
| | HMA modulus (MPa) | 2140 | 3774 | 3774 | 2470 |
| 6 | Average Pavement Temperature (°C) | 28 | 24 | 24 | 26 |
| | HMA modulus (MPa) | 2470 | 4328 | 4328 | 3282 |
| 7 | Average Pavement Temperature (°C) | 26 | 23 | 23 | 25 |
| | HMA modulus (MPa) | 3282 | 4948 | 4948 | 3774 |
| 8 | Average Pavement Temperature (°C) | 25 | 22 | 21 | 24 |
| | HMA modulus (MPa) | 3774 | 5636 | 6390 | 4328 |
| 9 | Average Pavement Temperature (°C) | 24 | 20 | 20 | 22 |
| | HMA modulus (MPa) | 4328 | 7211 | 7211 | 5636 |
| 10 | Average Pavement Temperature (°C) | 22 | 19 | 18 | 21 |
| | HMA modulus (MPa) | 5636 | 8093 | 9034 | 6390 |

**Table 4.** Regression analysis

| Aircraft | $\beta_0$ | $\beta_1$ | $R^2$ |
|----------|-----------|-----------|-------|
| B738 | 0.00209 | -0.31399 | 0.997 |
| A320 | 0.00183 | -0.31498 | 0.997 |
| A319 | 0.00196 | -0.31428 | 0.997 |
| B737 | 0.00207 | -0.32418 | 0.997 |
| E195 | 0.001653 | -0.33706 | 0.996 |

**Table 5.** Average damage per season using MCS.

| Aircraft | Damage per season | | | | CDF MCS | CDF FAARFIELD |
|----------|--------|------|--------|--------|---------|---------------|
| | **Summer** | **Fall** | **Winter** | **Spring** | | |
| **B738** | 0.028 | 0.016 | 0.019 | 0.027 | 0.090 | 0.430 |
| **B737** | 0.009 | 0.005 | 0.006 | 0.009 | 0.029 | 0.140 |
| **A320** | 0.012 | 0.007 | 0.008 | 0.012 | 0.039 | 0.180 |
| **A319** | 0.016 | 0.009 | 0.010 | 0.015 | 0.050 | 0.240 |
| **E195** | 0.001 | 0.000 | 0.000 | 0.001 | 0.002 | 0.010 |
| **Total** | 0.066 | 0.037 | 0.043 | 0.064 | 0.210 | 1.000 |

**FIGURE CAPTION LIST:**

**Fig. 1.** Procedure framework of this study.

**Fig. 2.** Annual operations at the airport.

**Fig. 3.** Air temperature variation over time.

**Fig. 4.** Air temperatures throughout the seasons in the Southern Hemisphere.

**Fig. 5.** Kolmogorov-Smirnov test for air temperature data across seasons In the Southern Hemisphere.

**Fig. 6.** Influence of pavement temperature on tensile strains in the asphalt layer.

**Fig. 7.** Influence of pavement temperature on compressive strains in the subgrade.

**Fig. 8a.** Convergence analysis for the MCS.

**Fig. 8b.** Standard deviation for the MCS.

**Fig. 9.** Reliability and design period extension.

**FIGURE FILES:**

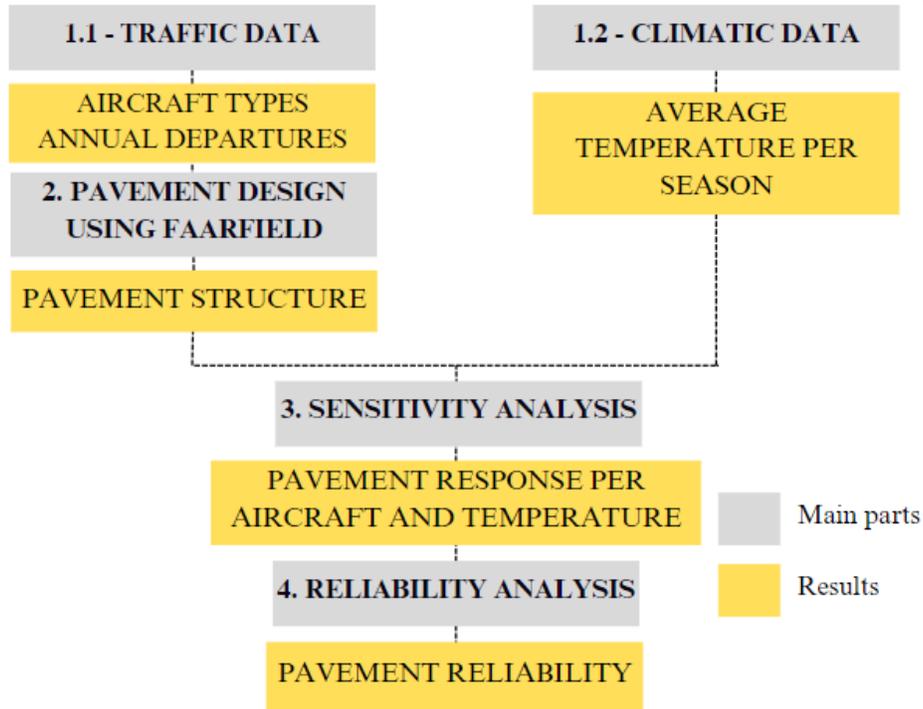

**Fig. 1.** Procedure framework of this study.

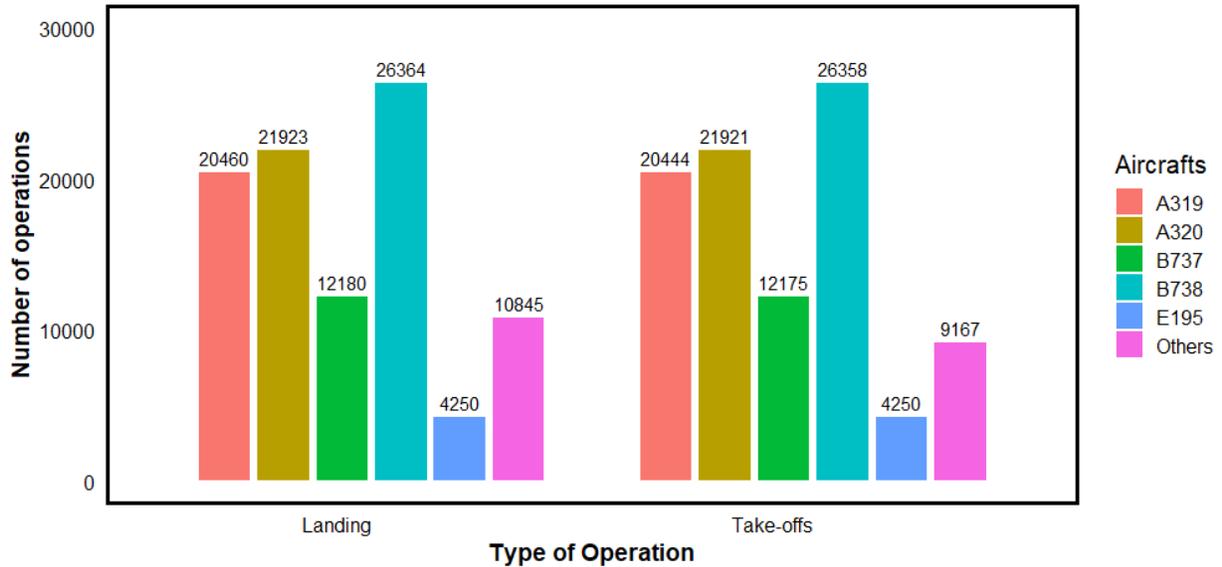

**Fig. 2.** Annual operations at the airport.

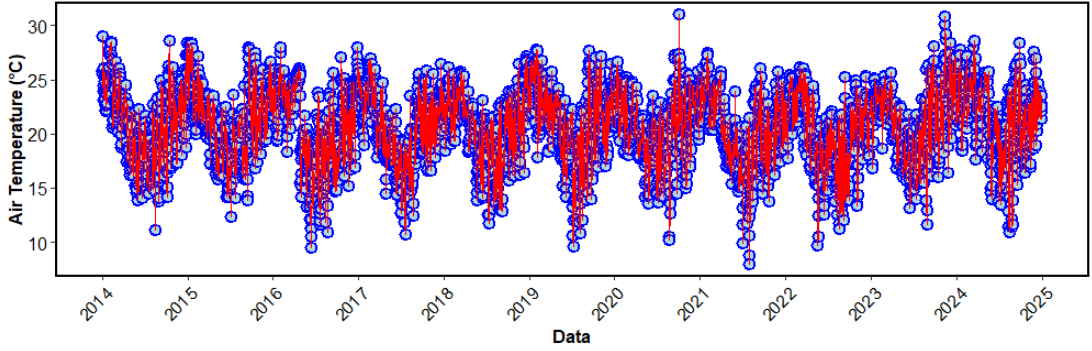

**Fig. 3.** Air temperature variation over time

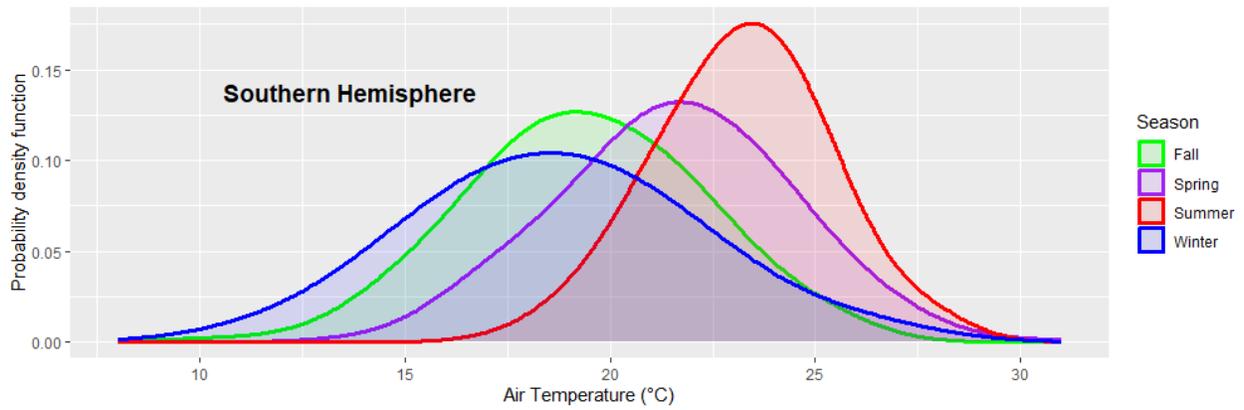

**Fig. 4.** Air temperatures throughout the seasons in the Southern Hemisphere

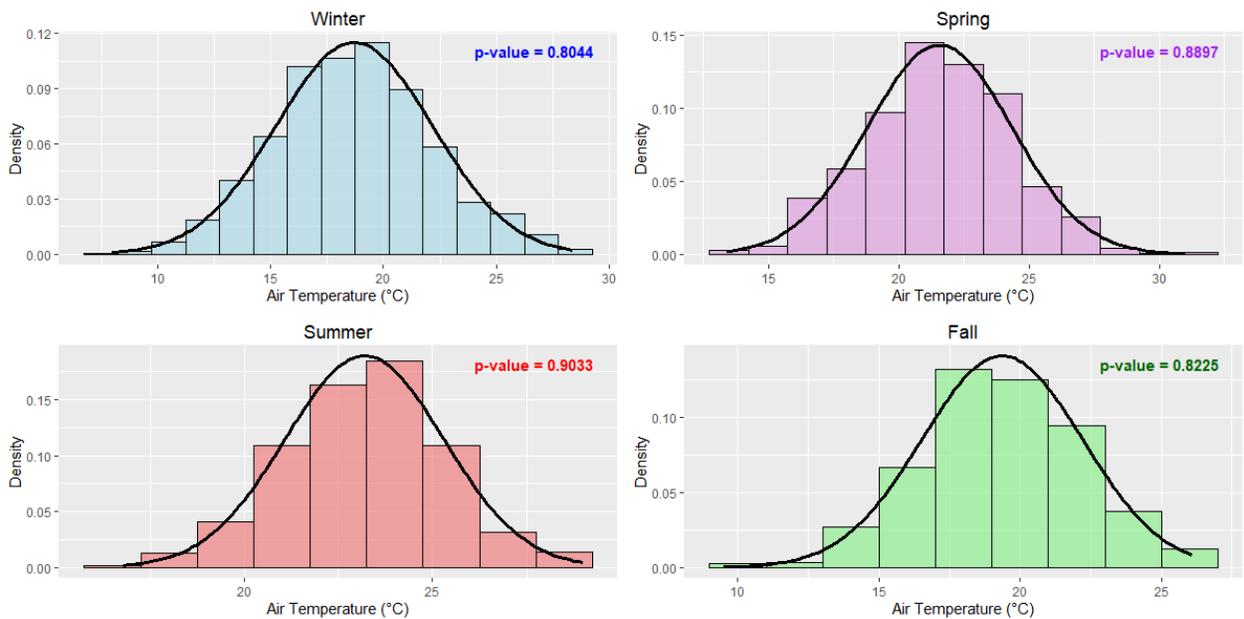

**Fig. 5.** Kolmogorov-Smirnov test for air temperature data across seasons in the Southern Hemisphere

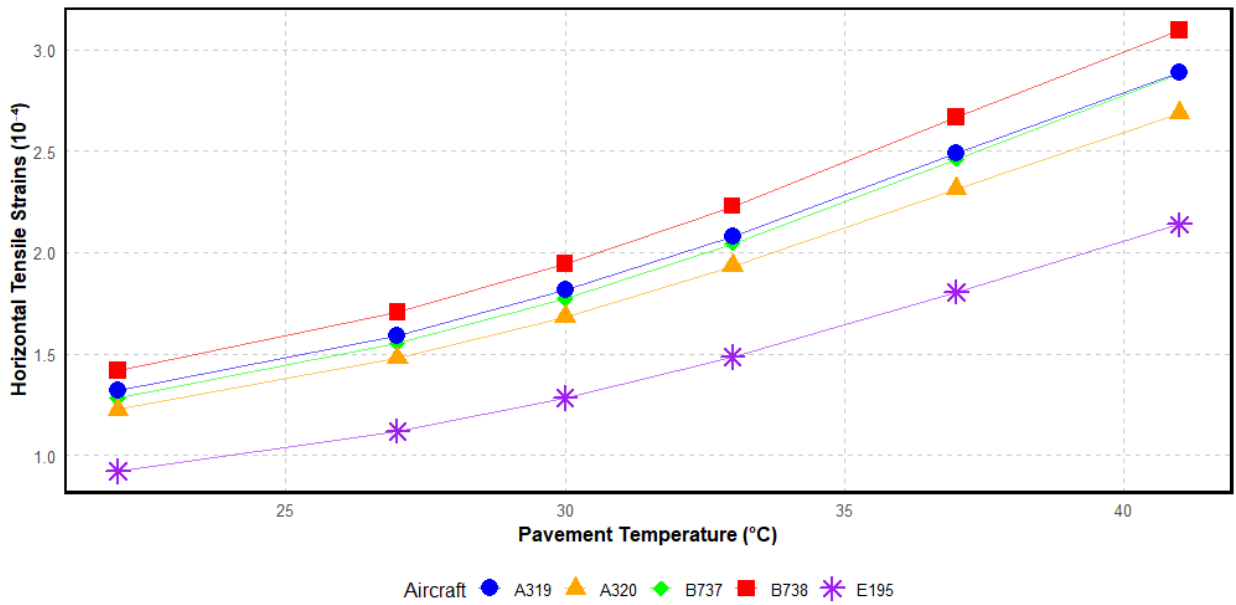

**Fig. 6.** Influence of pavement temperature on tensile strains in the asphalt layer.

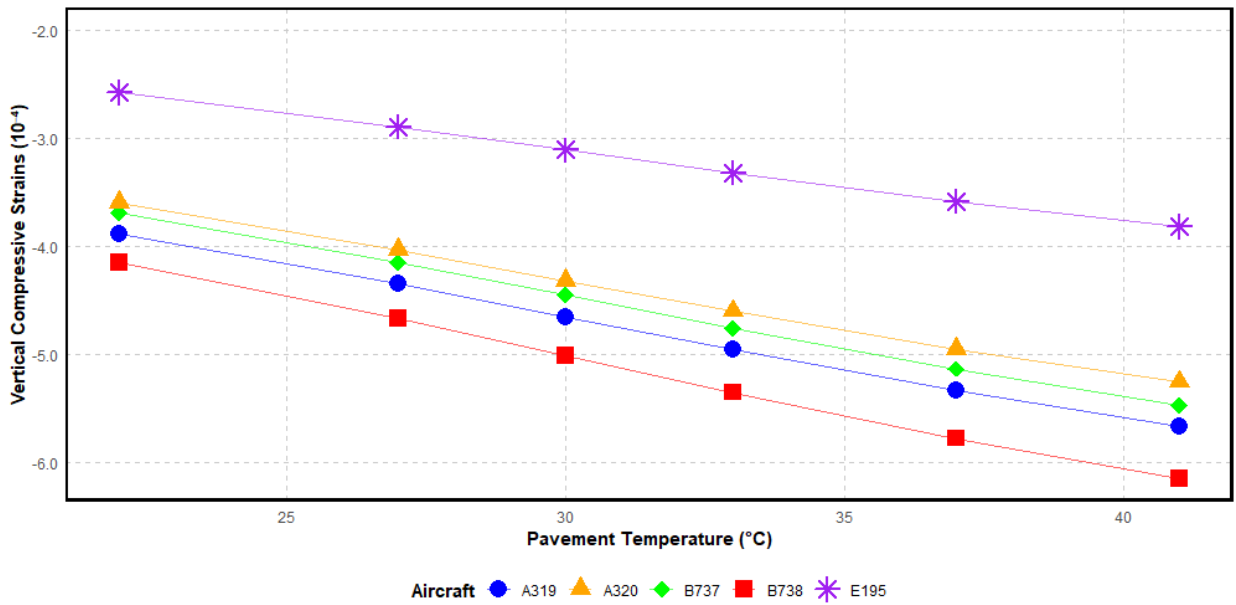

**Fig. 7.** Influence of pavement temperature on compressive strains in the subgrade

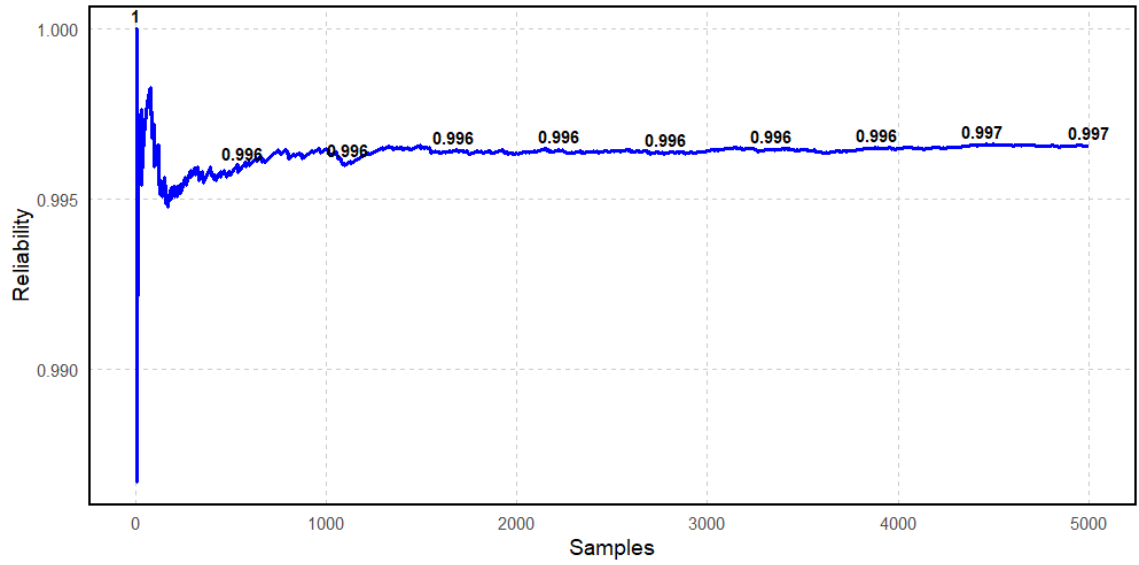

(a)

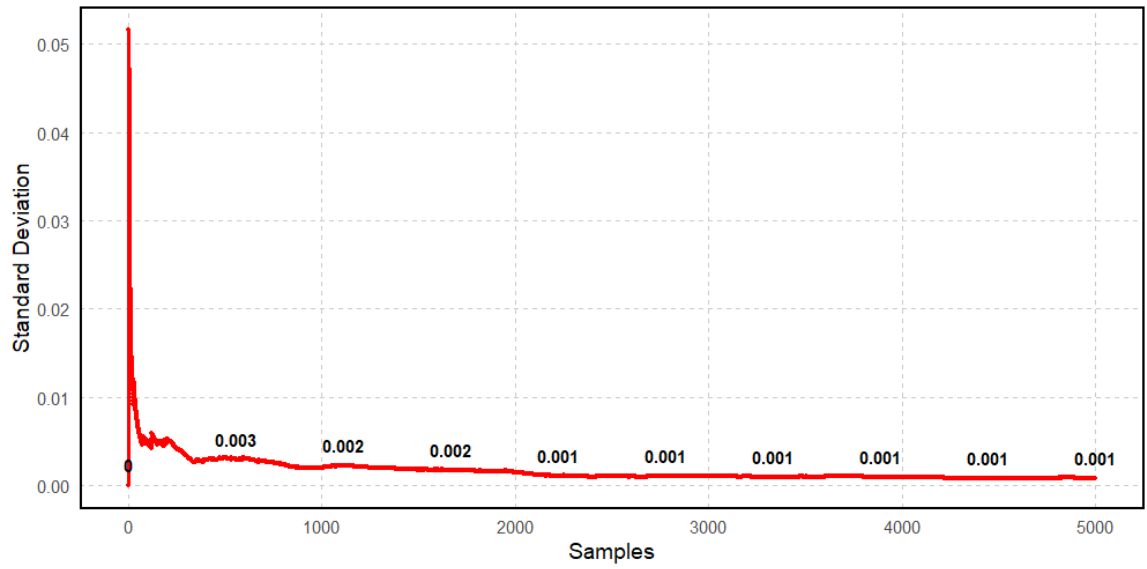

(b)

**Fig. 8.** Convergence analysis (a) and standard deviation (b) for the MCS

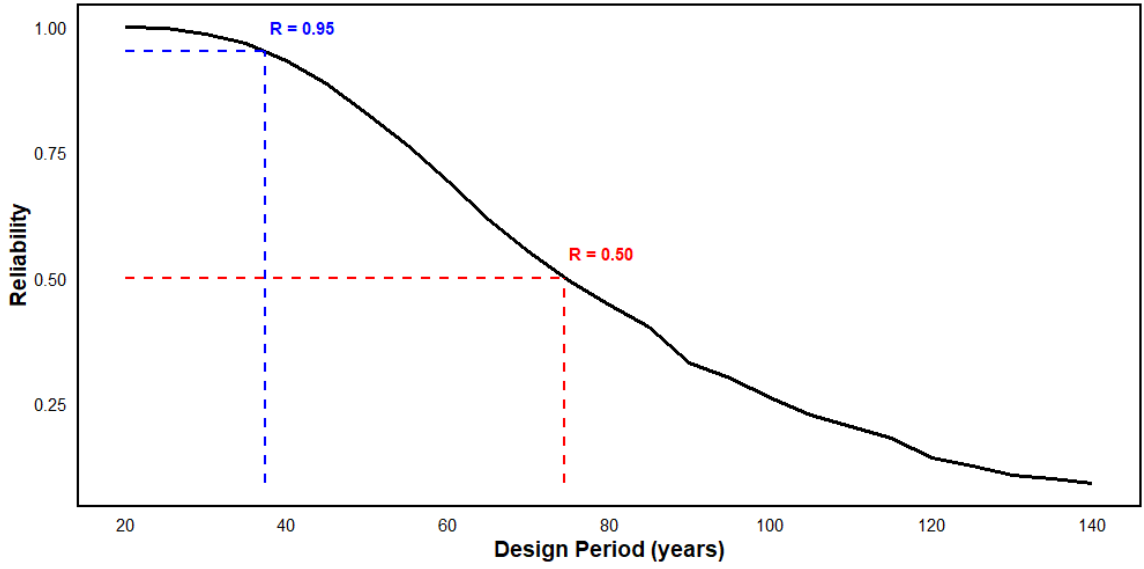

**Fig. 9.** Reliability and design period extension